\documentclass[twocolumn,showpacs]{revtex4}
\usepackage{graphicx}
\usepackage{dcolumn}
\usepackage{amsmath}
\usepackage{amssymb}
\begin{document}
\title{Search for $\bar{\nu}_e$ from the Sun at Super-Kamiokande-I}
\newcounter{foots}
\newcounter{notes}
\newcommand{\authoraticrr}{$^{a}$}
\newcommand{\authoratbu}{$^{b}$}
\newcommand{\authoratbnl}{$^{c}$}
\newcommand{\authoratuci}{$^{d}$}
\newcommand{\authoratcsu}{$^{e}$}
\newcommand{\authoratgmu}{$^{f}$}
\newcommand{\authoratgifu}{$^{g}$}
\newcommand{\authoratuh}{$^{h}$}
\newcommand{\authoratkek}{$^{i}$}
\newcommand{\authoratkobe}{$^{j}$}
\newcommand{\authoratkyoto}{$^{k}$}
\newcommand{\authoratlanl}{$^{l}$}
\newcommand{\authoratlsu}{$^{m}$}
\newcommand{\authoratumd}{$^{n}$}
\newcommand{\authoratmit}{$^{o}$}
\newcommand{\authoratduluth}{$^{p}$}
\newcommand{\authoratsuny}{$^{q}$}
\newcommand{\authoratnagoya}{$^{r}$}
\newcommand{\authoratniigata}{$^{s}$}
\newcommand{\authoratosaka}{$^{t}$}
\newcommand{\authoratseoul}{$^{u}$}
\newcommand{\authoratshizuokaseika}{$^{v}$}
\newcommand{\authoratshizuoka}{$^{w}$}
\newcommand{\authorattohoku}{$^{x}$}
\newcommand{\authorattokyo}{$^{y}$}
\newcommand{\authorattokai}{$^{z}$}
\newcommand{\authorattit}{$^{aa}$}
\newcommand{\authoratwarsaw}{$^{bb}$}
\newcommand{\authoratuw}{$^{cc}$}

\newcommand{\addressoficrr}[1]{$^{a}$ #1 }
\newcommand{\addressofbu}[1]{$^{b}$ #1 }
\newcommand{\addressofbnl}[1]{$^{c}$ #1 }
\newcommand{\addressofuci}[1]{$^{d}$ #1 }
\newcommand{\addressofcsu}[1]{$^{e}$ #1 }
\newcommand{\addressofgmu}[1]{$^{f}$ #1 }
\newcommand{\addressofgifu}[1]{$^{g}$ #1 }
\newcommand{\addressofuh}[1]{$^{h}$ #1 }
\newcommand{\addressofkek}[1]{$^{i}$ #1 }
\newcommand{\addressofkobe}[1]{$^{j}$ #1 }
\newcommand{\addressofkyoto}[1]{$^{k}$ #1 }
\newcommand{\addressoflanl}[1]{$^{l}$ #1 }
\newcommand{\addressoflsu}[1]{$^{m}$ #1 }
\newcommand{\addressofumd}[1]{$^{n}$ #1 }
\newcommand{\addressofmit}[1]{$^{o}$ #1 }
\newcommand{\addressofduluth}[1]{$^{p}$ #1 }
\newcommand{\addressofsuny}[1]{$^{q}$ #1 }
\newcommand{\addressofnagoya}[1]{$^{r}$ #1 }
\newcommand{\addressofniigata}[1]{$^{s}$ #1 }
\newcommand{\addressofosaka}[1]{$^{t}$ #1 }
\newcommand{\addressofseoul}[1]{$^{u}$ #1 }
\newcommand{\addressofshizuokaseika}[1]{$^{v}$ #1 }
\newcommand{\addressofshizuoka}[1]{$^{w}$ #1 }
\newcommand{\addressoftohoku}[1]{$^{x}$ #1 }
\newcommand{\addressoftokyo}[1]{$^{y}$ #1 }
\newcommand{\addressoftokai}[1]{$^{z}$ #1 }
\newcommand{\addressoftit}[1]{$^{aa}$ #1 }
\newcommand{\addressofwarsaw}[1]{$^{bb}$ #1 }
\newcommand{\addressofuw}[1]{$^{cc}$ #1 }

\author{
{\large The Super-Kamiokande Collaboration} \\
\bigskip
%
Y.~Gando\authorattohoku,
%
S.~Fukuda\authoraticrr,
Y.~Fukuda\authoraticrr,
M.~Ishitsuka\authoraticrr,
Y.~Itow\authoraticrr,
T.~Kajita\authoraticrr,
J.~Kameda$^{a,i}$,
K.~Kaneyuki\authoraticrr,
K.~Kobayashi\authoraticrr,
Y.~Koshio\authoraticrr
% ...
\\
\smallskip
\smallskip
\footnotesize
\it
\addressoficrr{Institute for Cosmic Ray Research, University of Tokyo, Kashiwa, Chiba 277-8582, Japan}\\
\addressofbu{Department of Physics, Boston University, Boston, MA 02215, USA}\\
\addressofbnl{Physics Department, Brookhaven National Laboratory, Upton, NY 11973, USA}\\
\addressofuci{Department of Physics and Astronomy, University of California, Irvine, Irvine, CA 92697-4575, USA }\\
\addressofcsu{Department of Physics, California State University, Dominguez Hills, Carson, CA 90747, USA}\\
\addressofgmu{Department of Physics, George Mason University, Fairfax, VA 22030, USA }\\
\addressofgifu{Department of Physics, Gifu University, Gifu, Gifu 501-1193, Japan}\\
\addressofuh{Department of Physics and Astronomy, University of Hawaii, Honolulu, HI 96822, USA}\\
\addressofkek{Institute of Particle and Nuclear Studies, High Energy Accelerator Research Organization (KEK), Tsukuba, Ibaraki 305-0801, Japan }\\
\addressofkobe{Department of Physics, Kobe University, Kobe, Hyogo 657-8501, Japan}\\
\addressofkyoto{Department of Physics, Kyoto University, Kyoto 606-8502, Japan}\\
\addressoflanl{Physics Division, P-23, Los Alamos National Laboratory, Los Alamos, NM 87544, USA }\\
\addressoflsu{Department of Physics and Astronomy, Louisiana State University, Baton Rouge, LA 70803, USA }\\
\addressofumd{Department of Physics, University of Maryland, College Park, MD 20742, USA }\\
\addressofmit{Department of Physics, Massachusetts Institute of Technology, Cambridge, MA 02139, USA}\\
\addressofduluth{Department of Physics, University of Minnesota, Duluth, MN 55812-2496, USA}\\
\addressofsuny{Department of Physics and Astronomy, State University of New York, Stony Brook, NY 11794-3800, USA}\\
\addressofnagoya{Department of Physics, Nagoya University, Nagoya, Aichi 464-8602, Japan}\\
\addressofniigata{Department of Physics, Niigata University, Niigata, Niigata 950-2181, Japan }\\
\addressofosaka{Department of Physics, Osaka University, Toyonaka, Osaka 560-0043, Japan}\\
\addressofseoul{Department of Physics, Seoul National University, Seoul 151-742, Korea}\\
\addressofshizuokaseika{Internatinal and Cultural Studies, Shizuoka Seika College, Yaizu, Shizuoka, 425-8611, Japan}\\
\addressofshizuoka{Department of Systems Engineering, Shizuoka University, Hamamatsu, Shizuoka 432-8561, Japan}\\
\addressoftohoku{Research Center for Neutrino Science, Tohoku University, Sendai, Miyagi 980-8578, Japan}\\
\addressoftokyo{The University of Tokyo, Tokyo 113-0033, Japan }\\
\addressoftokai{Department of Physics, Tokai University, Hiratsuka, Kanagawa 259-1292, Japan}\\
\addressoftit{Department of Physics, Tokyo Institute for Technology, Meguro, Tokyo 152-8551, Japan }\\
\addressofwarsaw{Institute of Experimental Physics, Warsaw University, 00-681 Warsaw, Poland }\\
\addressofuw{Department of Physics, University of Washington, Seattle, WA 98195-1560, USA}\\
}
\affiliation{ } 

\begin{abstract}
We present the results of a search for low energy $\bar{\nu}_e$ 
from the Sun using  1496 days of data from Super-Kamiokande-I.
We observe no significant excess of events and set an 
upper limit for the conversion probability to $\bar{\nu}_e$ of the
 $^8$B solar neutrino. This conversion limit is 0.8\% (90\% C.L.) of the
standard solar model's neutrino flux 
for total energy = 8~MeV - 20~MeV.
We also set a flux limit for monochromatic 
$\bar{\nu}_e$ for $E_{\bar{\nu}_e}$ = 10MeV - 17MeV.
\end{abstract}

%

\maketitle
\clearpage
Solar neutrino measurements at Super-Kamiokande~\cite{SK1500} 
and SNO~\cite{SNOCC} have established that the solar neutrino 
problem is explained by the transformation of electron 
neutrinos to other active neutrinos.
The mechanism for this transformation is generally assumed to be 
via neutrino flavor oscillations from $\nu_e$ to 
some superposition of $\nu_{\mu}$ and $\nu_{\tau}$.
However, measurements reported so far do not rule
out the possibility of spin flavor precession(SFP) in which 
some of the $\nu_e$ transform to antiparticles 
($\bar{\nu}_{\mu}$, $\bar{\nu}_{\tau}$).
In the so-called ``hybrid models''~\cite{VVO}, SFP 
and oscillation can transform solar neutrinos to $\bar{\nu}_e$  
if the neutrino is Majorana, it has a large magnetic moment, and 
the Sun has a large magnetic field.
If the neutrino has a magnetic moment, there are two 
possibilities: (1) the neutrino is a Dirac particle ; 
(2) it is a Majorana particle. 
In the Dirac neutrino case, $\nu_e^L$ changes to $\nu_e^R$ by 
the spin magnetic moment transition.
The $\nu_e^R$ is a sterile neutrino. 
On the other hand, in the Majorana neutrino scenario, 
SFP causes $\nu_e \to \bar{\nu}_{\mu,\tau}$. 
Neutrino oscillation then yields $\bar{\nu}_{\mu,\tau} \to \bar{\nu}_e$. 
Solar $\bar{\nu}_e$ could also originate from neutrino decay~\cite{decay}.
In this paper, we present a search for $\bar{\nu}_e$ from the Sun.

The inverse beta decay process, $\bar{\nu}_e + p \to n + e^+$, is 
predominant for $\bar{\nu}_e$ interactions in Super-Kamiokande (SK). 
The cross section for this process is two orders of magnitude 
greater than that for elastic scattering, and therefore SK has 
good sensitivity for the detection of solar $\bar{\nu}_e$. 
The positron energy is related to the neutrino energy by 
$E_{e^+} \approx E_{\bar{\nu}_e} - 1.3$~MeV.
The positron angular distribution relative to the incident 
$\bar{\nu}_e$ direction is nearly flat with a small energy 
dependent slope~\cite{Vogel}, which is in contrast 
to the sharply forward peaked elastic scattering distribution.
The difference between these distributions can be used to separate 
solar neutrino events from $\bar{\nu}_e$ events.

\begin{figure}[h]
\begin{center}
\includegraphics[scale=0.38]{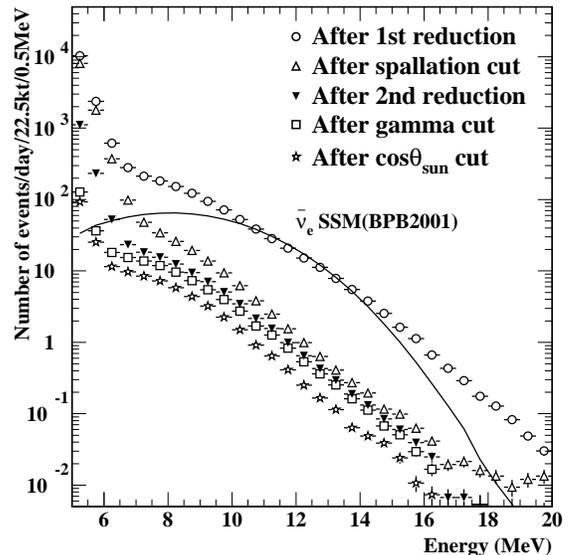}
\end{center}
 \vspace{-25pt}
 \caption{Energy spectrum after each reduction step. 
	  The solid curve shows the expected positron spectrum, 
	  after all cuts, assuming all $^8$B solar neutrinos 
	  convert to $\bar{\nu}_e$.
	  The horizontal axis shows the reconstructed total 
	  $e^{\pm}$ energy.}
\label{redstep}
\end{figure}

Super-Kamiokande is a 22.5 kton fiducial volume water Cherenkov detector, 
located in the Kamioka mine in Gifu, Japan. 
The data used for the search were collected in 1496 live days 
between May 31, 1996 and July 15, 2001. 
A detailed description of SK can be found elsewhere~\cite{SK1500,SK}.
Dominant backgrounds to the solar neutrino signal are $^{222}$Rn 
in the water, external gamma rays and muon-induced 
spallation products.
Background reduction is carried out in the following steps: 
first reduction, spallation cut, second reduction, and external 
$\gamma$-ray cut.
The first reduction removes events from electronic noise and 
other non-physical sources, and events with poorly reconstructed vertices.
The spallation cut removes events due to radio-isotopes (X) 
produced by cosmic ray muon interactions with water: 
$\mu + ^{16}O \to \mu + X$. 
These radio-isotopes are called ``spallation products.'' 
The spallation products emit beta and gamma rays and have lifetimes 
ranging from 0.001 to 14 sec. 
We cut these events using likelihood functions based on time, 
position, and muon pulse height.  The time and position 
likelihood functions are measures of the proximity of a candidate 
event to a muon track, while the pulse height likelihood
function measures the likelihood that a muon produced a shower.
These three likelihood functions are used together to
discriminate against spallation events~\cite{SK}.
The second reduction removes events with poor vertex fit quality 
and diffuse Cherenkov ring patterns, both characteristics of 
low-energy background events. 
The external $\gamma$-ray cut removes events due to $\gamma$-rays 
from the surrounding rock, photomultipliers(PMTs), etc.. Fig.~\ref{redstep} shows 
the energy spectrum after each reduction step.

At SK, a positron from inverse beta decay is  
indistinguishable from an electron or a gamma ray because 
the delayed 2.2 MeV gamma ray from n + p $\to$ d + $\gamma$ 
is below the detector's energy threshold.  
In order to remove elastic scattering events due to solar neutrinos, 
we cut events with cos$\theta_{\rm sun}$ $\ge$ 0.5, where 
$\theta_{\rm sun}$ is the event direction with respect to the 
direction from the Sun.
The region $\cos\theta_{\rm sun} <$ 0.5 would be occupied by 
solar $\bar{\nu}_e$ events, in addition to events due to 
known background sources which could not be removed by 
the standard data reduction. 
For E $\lesssim$~8~MeV, most background events are due to
radioactivity in the detector materials (such as $^{222}$Rn). 
Spallation accounts for a small fraction of background events 
in this region. 
In contrast, for E $\gtrsim$~8~MeV, most background events are 
produced by spallation.

\begin{figure}[h]
\begin{center}
\includegraphics[scale=0.43]{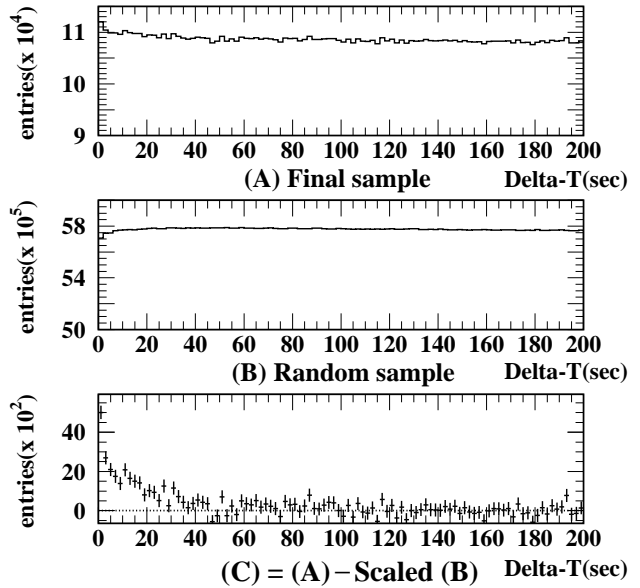}
\end{center}
 \vspace{-15pt}
 \caption{(A): $\mu$ delta-T distribution before observed events 
(B): Before random events 
(C): The delta-T distribution of events caused by spallation products 
obtained as (A) $-$ scale factor $\times$ (B).}
\label{dlnolimit}
\end{figure}

\begin{figure}[h]
\begin{center}
\includegraphics[scale=0.37]{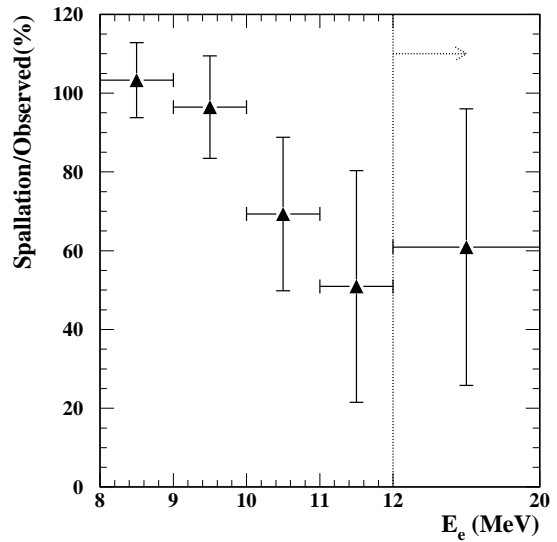}
\end{center}
 \vspace{-15pt}
 \caption{Spallation contamination in each energy bin. 
 The horizontal axis shows the total energy and the vertical 
 axis shows the ratio of spallation events to observed events.
 }
\label{spa_rate}
\end{figure}

The spallation cut used in the data reduction efficiently removes 
short-lifetime spallation products. This cut also removes $\sim$90\% 
of long-lifetime products such as $^{16}_{7}$N 
($\tau_{\frac{1}{2}}$ = 7.1 sec) and 
$^{11}_{4}$Be ($\tau_{\frac{1}{2}}$ = 13.8 sec).
Event by event removal of the remaining $\sim$ 10\% of these events 
is impractical because this introduces large dead time. 
However, we can estimate the contribution of these events to 
the post-reduction data sample using a statistical subtraction technique.
First, we made a time distribution of muon events 
preceding each low energy event by up to 200 seconds 
(Fig.~\ref{dlnolimit}(A)).
Since the average muon rate at SK is $\approx$ 2.5 Hz, there are, 
on average, $\approx$ 500 events for each low energy event. 
If the low energy event is due to a long lifetime spallation product, 
its event time will be correlated with one of the $\sim$500 preceding 
muon events.
If this is not the case, then its event time will be uncorrelated 
with all of the muon events.
To estimate the number of $\mu$ responsible for spallation events, 
we have to subtract the number of $\mu$ which did not make 
spallation events from the total number of $\mu$. 
In order to perform this subtraction, we made a sample of simulated 
events distributed randomly in space and time.
We applied the spallation cut to this sample 
as in the actual data sample in order to account for biases 
introduced by this cut. 
The muon time distribution of the random sample is shown in 
Fig.~\ref{dlnolimit}(B). The dip near Delta-T = 0 is due to 
the accidental loss of events by the spallation cut.
To estimate the number of muons which made spallation products, 
distribution (B) with suitable normalization is subtracted from 
distribution (A); the result of this subtraction is shown in 
Fig.~\ref{dlnolimit}(C).
The number of muon events in the delta-T = 100 sec - 200 sec 
region is used to calculate the normalization factor 
because the contamination from muons which make 
spallation products is negligible in this region.
The number of spallation events is obtained as
$$Spa = N_{0-50sec}^{observed} - N_{0-50sec}^{random} \times \frac{N_{100-200sec}^{observed}}{N_{100-200sec}^{random}} $$
$N_{0-50sec}^{observed}$ is the number of muon events within 50 seconds 
preceding the observed events, while $N_{0-50sec}^{random}$ is the 
corresponding number for random events. $N_{100-200sec}^{observed}$ 
and $N_{100-200sec}^{random}$ are similarly defined, but with a 
timing window of 100 to 200 seconds preceding the events.
For 8.0-20.0 MeV, and $\cos\theta_{\rm sun} \le 0.5$,
the number of spallation background events 
obtained by this method is (2.77 $\pm$ 0.20) $\times$ 10$^4$.
The number of observed $\bar{\nu}_e$ candidate events is 29781, 
so the ratio of spallation events 
to observed events is (93$\pm$7)\%.
The spallation contamination in each energy bin is 
shown in Fig.~\ref{spa_rate}.

The energy spectrum of the solar $\bar{\nu}_e$ is not known 
because the mechanism for $\bar{\nu}_e$ creation is not known.
Even if one assumes the SFP-oscillation hybrid model, 
the energy spectrum depends on $\mu_{\nu} \times$ B$_{\mbox solar}$,
$\Delta m^2$ and $\sin^2(2\theta)$, none of which are known precisely, 
if at all.
In order to deal with this ambiguity, we have chosen two spectrum 
models: the $^{8}$B neutrino spectrum~\cite{ortiz} and 
monochromatic spectrum (spectrum independent analysis).

For the $^8$B spectrum dependent analysis, we obtain an upper limit on
the solar $\bar{\nu}_e$ flux by comparing the observed number of events
outside of the elastic scattering peak ($\cos\theta_{\rm sun} \le 0.5$) 
with the expected number of $\bar{\nu}_e$ events assuming that all 
$^8$B neutrinos convert to $\bar{\nu}_e$. 
The expected number is obtained by Monte Carlo
simulation of solar $\bar{\nu}_e$ interaction with the detector. 
The $\cos\theta_{\rm sun}$ dependence was simulated , and 
the effect of this dependence on the $\bar{\nu}_e$ efficiency is 
taken into account.
The standard solar model(SSM) $^8$B neutrino flux was assumed
(5.05 $\times 10^6$ /cm$^{2}$/sec)~\cite{BPB2001} . 
Through the rest of this paper, electron neutrino spectrum and 
flux refer to the unoscillated quantities at the Sun. 
The solid lines in Fig.4 show 90\% C.L. limits on the $\bar{\nu}_e$ 
flux before statistical spallation subtraction. 
The dashed lines show the limits after statistical subtraction 
(only for E $\ge$ 8 MeV). 
By combining the statistics for 8 MeV $\le$ E $\le$ 20 MeV, 
we obtained a global upper limit of 0.8\% of the SSM neutrino flux.

Some authors have indicated that the positron angular 
distribution may be useful for the search for $\bar{\nu}_e$ 
in the SK data (e.g. \cite{ang0,Lujan}). 
$\cos\theta_{\rm sun}$ is distributed as 
f($\cos\theta_{\rm sun}$) = 0.5 $\times$ (1 + 
$\alpha \times \cos\theta_{\rm sun}$), where $\alpha$ 
is a monotonically increasing function of neutrino energy 
(except near threshold), and $\alpha$ $<$ 0 for $E_{\nu} \lesssim~13$~MeV 
and $>$ 0 above this~\cite{Vogel}.
The angular information is useful for the $\bar{\nu}_e$
search at the lowest neutrino energies where
f($\cos\theta_{\rm sun}$) has sufficient slope and the event
statistics are large.
$\bar{\nu}_e$ events with the predicted $\cos\theta_{\rm sun}$ 
distribution were input to a detector simulator to obtain 
the expected positron angular distribution.
The resulting distribution has the same form as above.
The fit value of $\alpha$ is $-$0.076 at $E$ = 5 - 6 MeV, 
0.107 at $E$ = 12 - 20 MeV, and crosses 0 at  $\sim$ 9 MeV.

Solar neutrino elastic scattering is one of the backgrounds 
for this analysis. 
Almost all such events have $\cos\theta_{\rm sun} >$ 0.5, 
so events with $\cos\theta_{\rm sun} >$ 0.5 are cut. 
We also subtract the small amount of spill-over into 
$\cos\theta_{\rm sun}$ $\le$ 0.5 using Monte Carlo simulation
($\sim$5\% for 5-20MeV). 
Another background is due to $^{18}$O($\nu_e$;$e$)$^{18}$F~\cite{o18}. 
There is only a small number of events from this source 
(0.03\% $\sim$ 2\%, depending on energy),
but electrons from this process, like the low-energy 
$\bar{\nu}_e$, have negative slope in their angular distribution.
So they are subtracted from the data.
The $\nu_e$ flux is taken as the charged current flux value from 
SNO, $1.76 \times 10^6$ /cm$^{2}$/sec ~\cite{SNO}.

\begin{figure}[h]
\begin{center}
\includegraphics[scale=0.41]{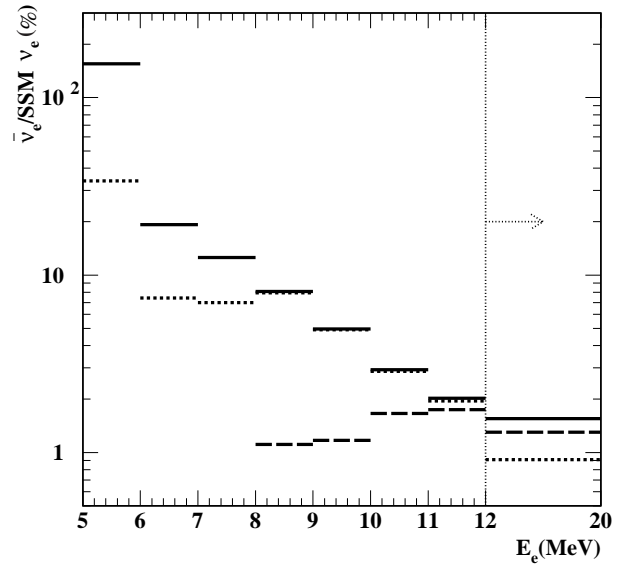}
\end{center}
 \caption{Summary of $\bar{\nu}_e$ limits.
 The horizontal axis shows total positron energy 
 and the vertical axis shows the 90\% C.L. $\bar{\nu}_e$ 
 rate normalized to the SSM $\nu_e$ rate.
 The solid lines show the 90\% C.L. limit ratio. 
 The dashed lines show the limit after statistical subtraction of 
 the spallation background. 
 The dotted lines show the result from the angular distribution analysis. }
\label{summary}
\end{figure}

A $\bar{\nu}_e$ flux upper limit is obtained using 
a probability test with the slope of the $\cos\theta_{\rm sun}$ 
distribution serving as a constraint.
This test is based on a $\chi^2$ test with $\chi^2(\delta,\beta,\gamma)$ 
defined for each energy as follows: \\
$$\sum_{i=1}^{N_{\cos}}\left\{\frac{N^{data}_i-N^{el}_i-N^{^{18}O}_i- \delta \cdot N^{\bar{\nu}_e}_i-\beta \cdot n^{BG}_i(1+\gamma \cdot x_i)}{\sigma^{stat.}_i}\right\}^2$$ $$ + \left(\frac{\gamma}{\sigma^{syst.}}\right)^2$$
$i$ is the index for the $\cos\theta_{\rm sun}$ bins 
($\cos\theta_{\rm sun} \le$ 0.5, $N_{\cos}$= 30), 
$x_i$ is $(\cos\theta_{\rm sun})_i$,
$N^{data}_i$ is the number of observed data events, 
$\sigma^{stat.}_i$ is the statistical error of the observed data, 
$N^{el}_i$ is the expected number of elastic scattering events, 
$N^{^{18}O}_i$ is the expected number of events from 
the $^{18}$O($\nu_e$;$e$)$^{18}$F reaction, 
$N^{\bar{\nu}_e}_i$ is the number of $\bar{\nu}_e$ events 
assuming all SSM $\nu_e$ convert to $\bar{\nu}_e$
(the number in each bin $i$ depends on the slope $\alpha$), 
and $n^{BG}_i$ is the shape for all other background events 
that are almost uncorrelated in direction with the Sun
(this background is essentially flat). 
$N^{el}_i$ and $N^{^{18}O}_i$ are both $\lesssim$ 2\% of $N^{data}_i$, 
and the systematic errors of these terms are negligible.
$\sigma^{syst.}$(= 0.5\%) is the systematic error of the
slope of the background shape and $\gamma$ is 
the parameter that takes this into account.
$\beta$ parameterizes the amount of such background events.
We divided the parameter space for $\delta$ into a grid, 
and minimized $\chi^2$ with respect to $\beta$ and $\gamma$ 
at each grid point. 
The resulting $\gamma$ and $\chi^2_{min}$  indicated good 
fits to the data. 
$\chi^2$ as a function $\delta$ obtained 
in this way is input to a probability function.
From this analysis, we set a 90\% C.L. upper limit for 
each energy bin.
The dotted lines in Fig.~\ref{summary} show the result. 
It should be noted that the spallation background
subtratction is not applied in this analysis for two
reasons.  First, for E $<$ 8 MeV, spallation subtraction
is ineffective because spallation events form a small
subset of the total background.  Second, for E $>$ 8
MeV, there are insufficient statistics after spallation
subtraction to perform an angular analysis of the
data.

\begin{figure}[h]
\begin{center}
\includegraphics[scale=0.41]{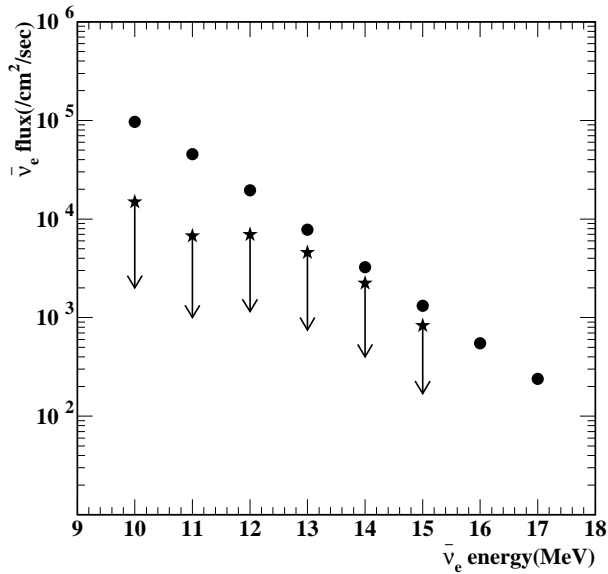}
\end{center}
 \caption{$\bar{\nu}_e$ flux 90\% C.L. upper limit for each 
 monochromatic $\bar{\nu}_e$.
 The horizontal axis shows neutrino energy and 
 the vertical axis shows the flux limit. 
 The circles show the limits before spallation 
 subtraction while the stars show the limits after subtraction. 
 The two highest-energy bins have insufficient number of events for 
 statistical subtraction.}
\label{spec_independ}
\end{figure}

The analysis described so far assume that the $\bar{\nu}_e$ originate from 
$^8$B solar neutrinos. 
We also generalized our search by assuming a monochromatic 
$\bar{\nu}_e$ source at various energies and set 
conservative $\bar{\nu}_e$ flux upper limits.
The interaction of such $\bar{\nu}_e$ with the detector was 
simulated, and standard data reduction cuts were applied. 
The positron spectrum is well described by a Gaussian. 
We then counted the number of events in the data in the 
$\pm$ 1$\sigma$ range of this Gaussian. 
We took this number to be the number of events due to 
monochromatic $\bar{\nu}_e$ and obtained an upper limit. 
This upper limit is very conservative because we do not take 
account of the large spill-over from lower energy bins that is 
implied by the sharply falling spectrum seen in the data. 
We also obtained limits after statistical subtraction of 
long lifetime spallation events. The 90\% C.L. limits are shown in
Fig\ref{spec_independ}.

In summary, a search for $\bar{\nu}_e$ flux from the Sun 
was performed using all 1496 live days of solar neutrino 
data from Super-Kamiokande-I.
Using the $^8$B and monochromatic energy spectra,
90\% C.L. upper limits were set for the $\bar{\nu}_e$ flux. 
For the $^8$B spectrum dependent analysis, the upper limit 
to the flux was 0.8\% of the SSM $\nu_e$ flux prediction 
for E = 8.0-20.0MeV.
This can be compared with the Kamiokande result of 4.5\%~\cite{KID}.
For $\bar{\nu}_e$ fluxes with  various monochromatic energies, the 
resulting upper limits are shown in Fig.~\ref{spec_independ}.

\begin{acknowledgments}
The authors acknowledge the cooperation of the Kamioka Mining and 
Smelting Company.
The Super-Kamiokande has been built and operated from funding 
by the Japanese Ministry of Education, Culture, Sports, Science and 
Technology, the U.S. Department of Energy, and 
the U.S. National Science Foundation. 
This work was partially supported by the Korean Research
Foundation (BK21) and the Korea Ministry of Science and
Technology.

\end{acknowledgments}

\end{document}